\documentclass[cameraready]{Interspeech}

\title{HRIR-Former: Grid-Free Time-Domain Reconstruction of Head-Related Impulse Responses with a Spatially Encoded Transformer}

\author[affiliation={1}, orcid=0009-0007-6962-4574, equalcontribution, correspondingauthor]{Shaoheng}{Xu}
\author[affiliation={1}, orcid=0009-0002-7144-1332, equalcontribution]{Chunyi}{Sun}
\author[affiliation={2,1}, orcid=0000-0001-6817-139X]{Jihui (Aimee)}{Zhang}
\author[affiliation={1}, orcid=0000-0003-4942-7526]{Amy}{Bastine}
\author[affiliation={1}, orcid=0000-0002-5589-4203]{Prasanga N.}{Samarasinghe}
\author[affiliation={1}, orcid=0000-0001-6937-7218]{Thushara D.}{Abhayapala}
\author[affiliation={1}, orcid=0000-0003-4125-1554]{Hongdong}{Li}

\address{
    $^1$ The Australian National University, Australia \\
    $^2$ The University of Queensland, Australia
}

\email{Shaoheng.Xu@anu.edu.au, Chunyi.Sun@anu.edu.au}


\usepackage{comment}

\usepackage{booktabs}
\usepackage{makecell}
\usepackage{array}
\usepackage{multirow}
\usepackage{eso-pic}

\begin{document}


\AddToShipoutPictureFG*{%
  \AtPageLowerLeft{%
    \put(
      \LenToUnit{\dimexpr 1in+\oddsidemargin\relax},
      \LenToUnit{3.5mm}
    ){%
      \makebox(0,0)[bl]{%
        \footnotesize
        \begin{minipage}[b]{\textwidth}
          \raggedright
          Published in Proc.\ Interspeech 2026, pp.\ 690--695. DOI: \url{https://doi.org/10.21437/Interspeech.2026-702}\\[1mm]
          Code: \url{https://github.com/ShaoHenry/HRIR-Former}\\[1mm]
          Demo: \url{https://shaohenry.github.io/HRIR-Former}
        \end{minipage}%
      }%
    }%
  }%
}

\maketitle
 
\begin{abstract}
    Individualized head-related impulse responses (HRIRs) enable binaural rendering, but dense per-listener measurements are costly. We address HRIR spatial up-sampling from sparse per-listener measurements: given a few measured HRIRs for a listener, predict HRIRs at unmeasured target directions. Prior learning methods often work in the frequency domain, rely on minimum-phase assumptions or separate timing models, and use a fixed direction grid, which can degrade temporal fidelity and spatial continuity. We propose HRIR-Former, a time-domain, grid-free binaural Transformer for reconstructing HRIRs at arbitrary directions from sparse inputs. It uses sinusoidal spatial features, a Conv1D refinement module, and auxiliary interaural time difference (ITD) and interaural level difference (ILD) heads. On SONICOM, it improves normalized mean squared error (NMSE), cosine distance, and ITD/ILD errors over prior methods; ablations validate modules and show minimum-phase preprocessing is unnecessary.
\end{abstract}


\section{Introduction}
\label{sec:introduction}

Spatial audio seeks to capture and reproduce realistic acoustic scenes by preserving both audio content and spatial cues, enabling an immersive listening experience~\cite{surroundbysound, review-cam}. Among spatial-audio techniques, binaural recording and rendering are particularly effective because they aim to reproduce the signals received at the listener’s two ears~\cite{surroundbysound, bin-1, bin-2}. Binaural rendering typically relies on head-related transfer functions (HRTFs) or their time-domain counterparts, head-related impulse responses (HRIRs), to reproduce direction-dependent spectral and timing cues that support human spatial hearing. As a result, binaural spatial audio has been widely adopted in virtual/augmented reality (VR/AR)~\cite{spatial2002, resonance, spatialaudiorender} and object-based audio production for film and other media~\cite{object-audio}. However, acquiring individualized HRTFs/HRIRs remains challenging: per-listener measurements are tedious, costly, and time-consuming, require specialized anechoic facilities, and may require listeners to remain motionless for around an hour~\cite{review-hrtf}. Accordingly, prior work has largely focused on two directions: \emph{HRTF personalization} and \emph{spatial up-sampling/augmentation}.

HRTF personalization estimates a listener’s HRTF from side information such as anthropometric measurements of the head, torso, and pinna~\cite{pHRTF, hscma-hrtf, survey-pHRTF}. Early approaches relied on numerical simulation and similarity-based selection~\cite{PRTFs}. More recent methods employ deep neural networks (DNNs) to learn non-linear mappings from anthropometric descriptors to HRTFs~\cite{dnn-1, dnn-2, dnn-3, dnn-4, dnn-5, dnn-6}, including latent-space approaches that encode HRTFs into a low-dimensional representation and then decode them back to full HRTFs~\cite{hrtf-latent}. However, simulation-based methods often require high-quality 3D geometry of the listener’s head and pinna~\cite{PRTFs}, while DNN-based approaches are constrained by the limited amount of available measured data~\cite{hrtf-latent}.

\looseness=-1 To reduce the time and practical complexity of HRTF measurements, spatial up-sampling and augmentation methods have been proposed. These methods aim to reconstruct a high angular-resolution HRTF set from only a small number of measurements at sparsely sampled directions~\cite{xie-HRTFbook}. Classical approaches typically represent HRTFs using a set of basis functions (BFs) and synthesize responses at unmeasured directions via interpolation or superposition~\cite{ana-1, ana-2}. In practice, these methods often degrade when measurements are spatially sparse (e.g., with angular gaps of $30$--$40^\circ$)~\cite{lap}. Spherical-harmonic (SH)-based interpolation is another common approach~\cite{sh1, sh-hrtf-2004, jasa-hrtf-sh, sh2, sh3, sh4, sh5}, but it can also perform poorly under sparse sampling~\cite{lap}. The Listener Acoustic Personalization (LAP) challenge~\cite{lap} has stimulated ML-based spatial augmentation, with results showing that learned models can outperform SH/BF interpolation under sparse sampling when reconstructing 793-direction HRTF sets from only 3--100 measurements~\cite{LAPreport, GEP-GAN, SYT-FSP-AE, ranf, hrtf-former, PINN-HRTF, sphericalCNN-HRTF}; however, most of these methods remain tied to reconstruction on a fixed grid of directions.

A further limitation of many existing ML-based pipelines is that they operate primarily in the frequency domain and focus on magnitude responses, since phase is difficult to model directly. Phase and timing cues are therefore often recovered using minimum-phase (MP) approximations and/or a separate estimate of the time-delay (time-of-arrival) term~\cite{ranf, pca-hrtf}. In contrast, relatively few works perform personalization or spatial up-sampling directly in the time domain. For example, \cite{icassp-hrir} proposes HRIR personalization using diffusion models guided by anthropometric inputs, but targets a fixed grid of source directions and does not support arbitrary directions. The time-domain decomposition and superposition methods in~\cite{fa-hrir-1,fa-hrir-2} enable spatially continuous HRIR reconstruction, but are restricted to the median plane. Thus, time-domain, spatially continuous HRIR reconstruction over arbitrary 3D directions remains under-explored.

Time-domain modeling is important for both room impulse responses (RIRs) and HRIRs, as phase and timing cues are central to spatial perception. Our prior work shows that transformer architectures with positional encoding can capture long-range temporal structure and enable continuous spatial modeling of RIRs~\cite{rir-former}. HRIR reconstruction, however, poses different challenges: the number of measured subjects is limited, inter-subject variability is substantial, and the model must preserve binaural cues such as interaural time difference (ITD) and interaural level difference (ILD). We therefore develop a framework tailored to continuous HRIR reconstruction over arbitrary 3D directions under realistic data constraints.

We propose \textit{HRIR-Former}, a time-domain, grid-free binaural Transformer for continuous angular (azimuth--elevation) HRIR spatial up-sampling that reconstructs HRIRs at arbitrary target directions from sparse per-listener measurements. The main contributions are:
(1) \textbf{Time-domain binaural reconstruction} that preserves phase information naturally and does not impose a MP assumption.
(2) \textbf{Grid-free directional modeling} via sinusoidal encoding of source positions, enabling inference at previously unseen target directions without requiring a fixed measurement grid.
(3) \textbf{Auxiliary binaural-cue supervision} using ITD/ILD-prediction heads to encourage accurate interaural timing and level cues.
(4) \textbf{Temporal refinement} via a post-Transformer Conv1D module to improve local directional consistency.
(5) \textbf{Time--frequency supervision} via time-domain losses and a complex HRTF loss $\mathcal{L}_{\text{HRTF}}$ for spectral alignment.


\section{Problem Formulation}
\label{sec:formulation}

\begin{figure}[t]
  \centering
  \includegraphics[width=0.7\linewidth]{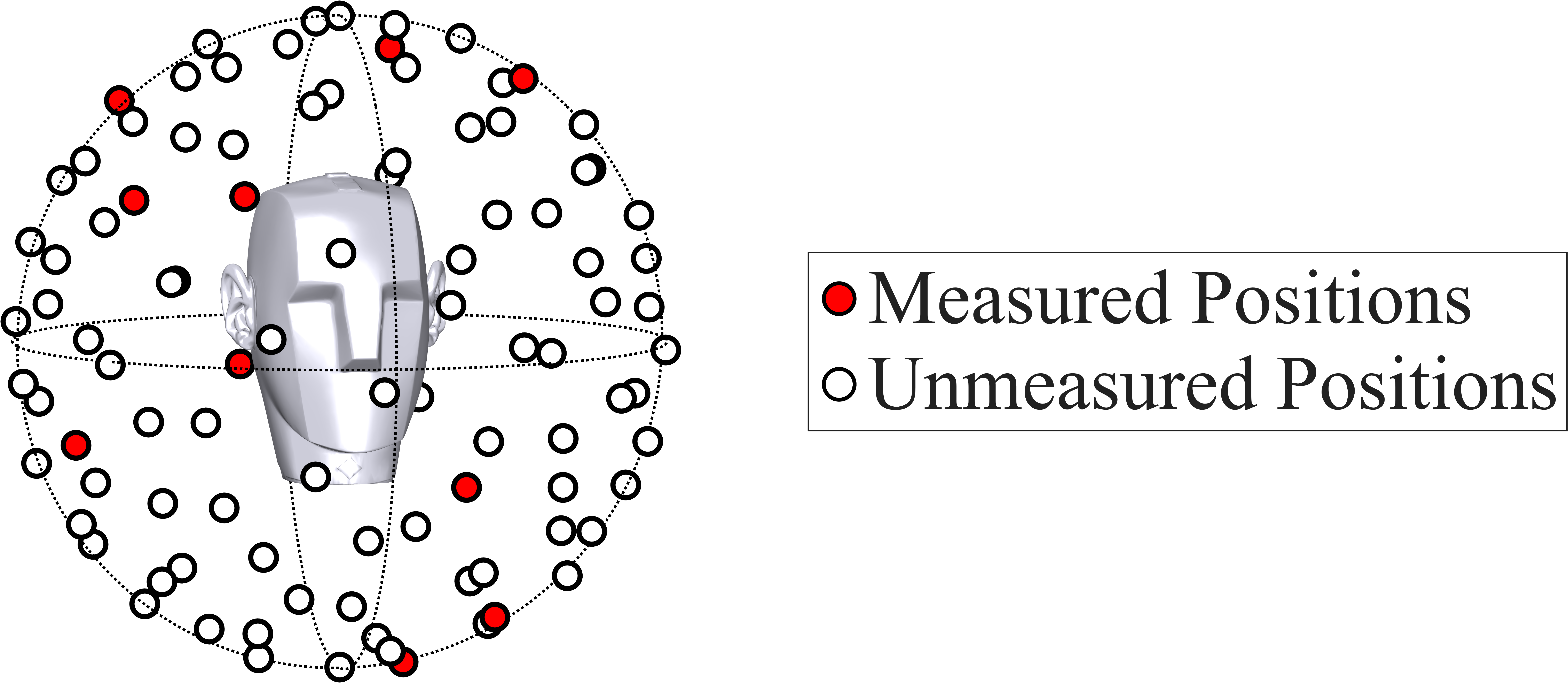}
  \caption{Illustration of HRIR spatial up-sampling: estimating HRIRs at unmeasured target directions (white) from a sparse set of measured directions (red).}
  \label{fig:setup}
  \vspace{-0.6cm}
\end{figure}

The objective of this work is to reconstruct HRIRs at arbitrary, unmeasured directions from a limited set of measured directions for a given subject. Consider a free-field HRIR measurement setup (without room reverberation). For each subject, let there be $L=M+N$ source directions $\mathbf{x}_\ell \triangleq (\phi_\ell,\,\theta_\ell,\,r_\ell)$ for $\ell=1,\dots,L$, where $\phi$ and $\theta$ denote azimuth and elevation (in degrees), and $r$ denotes the source radius (in meters) in a spherical coordinate system. Among these $L$ directions, $M$ are measured and $N$ are unmeasured target directions. The listener’s head is centered at the global origin $O$, as illustrated in Fig.~\ref{fig:setup}.

At each measured direction $\{\mathbf{x}^{\text{meas}}_m\}_{m=1}^{M}$, we record a binaural HRIR pair sampled at rate $f_s$ and of length $K$,
$\mathbf{h}^{(L)}_m,\,\mathbf{h}^{(R)}_m \in \mathbb{R}^{1\times K}$ for the left and right ears. We form a binaural HRIR row vector by concatenation
$\mathbf{h}^{\text{meas}}_m \triangleq \big[\mathbf{h}^{(L)}_m~\mathbf{h}^{(R)}_m\big] \in \mathbb{R}^{1\times 2K}$.
Stacking the $M$ measured binaural HRIRs yields
\begin{equation}
\label{eq:H}
\mathbf{H} =
\begin{bmatrix}
\mathbf{h}^{\text{meas}}_1 & \mathbf{h}^{\text{meas}}_2 & \cdots & \mathbf{h}^{\text{meas}}_M
\end{bmatrix}^{\top}
\in \mathbb{R}^{M \times 2K},
\end{equation}
where $(\cdot)^{\top}$ denotes transpose.

Our goal is to estimate binaural HRIRs at the $N$ target directions $\{\mathbf{x}^{\text{tgt}}_n\}_{n=1}^{N}$ at the same sampling rate $f_s$ and length $K$. Let $\bar{\mathbf{h}}^{(L)}_n, \bar{\mathbf{h}}^{(R)}_n$ and $\hat{\mathbf{h}}^{(L)}_n, \hat{\mathbf{h}}^{(R)}_n$ denote the ground-truth and estimated HRIRs for the two ears at direction $\mathbf{x}^{\text{tgt}}_n$. Define
\begin{equation}
\bar{\mathbf{h}}^{\text{tgt}}_n \triangleq \big[\bar{\mathbf{h}}^{(L)}_n~ \bar{\mathbf{h}}^{(R)}_n\big], \quad
\hat{\mathbf{h}}^{\text{tgt}}_n \triangleq \big[\hat{\mathbf{h}}^{(L)}_n~ \hat{\mathbf{h}}^{(R)}_n\big]
\in \mathbb{R}^{1\times 2K}.
\end{equation}
Stacking these yields $\bar{\mathbf{H}},\hat{\mathbf{H}}\in\mathbb{R}^{N\times 2K}$:
\begin{equation}
\label{eq:HbarHhat}
\bar{\mathbf{H}}=\big[\bar{\mathbf{h}}^{\text{tgt}}_1~\cdots~\bar{\mathbf{h}}^{\text{tgt}}_N\big]^{\top},\quad
\hat{\mathbf{H}}=\big[\hat{\mathbf{h}}^{\text{tgt}}_1~\cdots~\hat{\mathbf{h}}^{\text{tgt}}_N\big]^{\top}.
\end{equation}

The HRIR spatial up-sampling task can be formulated as learning a mapping
$\hat{\mathbf{H}}=\mathcal{F}(\mathbf{H},\{\mathbf{x}^{\text{meas}}_m\}_{m=1}^{M},\{\mathbf{x}^{\text{tgt}}_n\}_{n=1}^{N})$
such that $\hat{\mathbf{H}}$ closely matches $\bar{\mathbf{H}}$, i.e.,
\begin{equation}
\label{eq:objective}
\hat{\mathbf{H}} =
\arg\min_{\hat{\mathbf{H}}}
\left\lVert \bar{\mathbf{H}} - \hat{\mathbf{H}} \right\rVert_F^2
\quad \text{given} \quad \mathbf{H},\, \{\mathbf{x}^{\text{meas}}_m\},\, \{\mathbf{x}^{\text{tgt}}_n\},
\end{equation}
where $\lVert \cdot \rVert_F$ denotes the Frobenius norm.


\section{Proposed Method}
\label{sec:method}

\begin{figure*}[t!]
  \centering
  \includegraphics[width=1.0\linewidth]{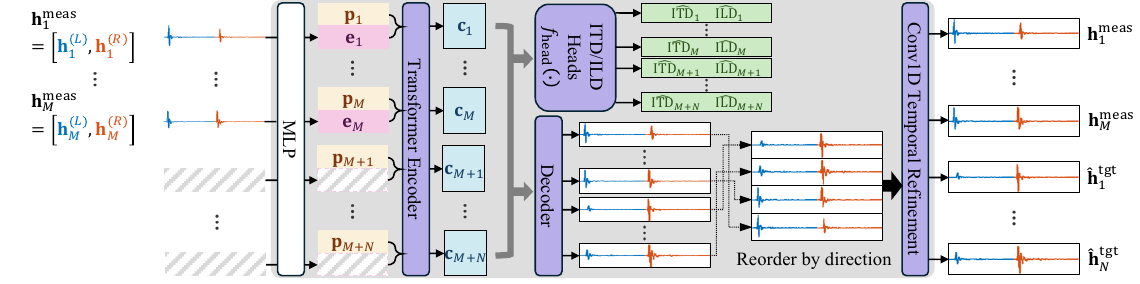}
  \vspace{-0.5cm}
  \caption{\textbf{Overview of the proposed \textit{HRIR-Former} pipeline.} Known HRIRs are projected into a latent space using an MLP-based signal encoder, while geometric embeddings derived from direction coordinates are independently projected and added to the signal features. The resulting tokens are processed jointly by a Transformer encoder to capture global spatial dependencies across measured and target directions. A shared MLP decoder maps contextual features to full-length binaural HRIRs for all directions, followed by masked fusion to preserve measured responses and a lightweight Conv1D for temporal refinement.}
  \label{fig:pipeline}
\vspace{-0.5cm}
\end{figure*}

We learn the mapping defined in Eq.~\ref{eq:objective} by casting HRIR spatial up-sampling as masked inpainting over the spherical direction set, where unmeasured directions are treated as masked tokens to be predicted from the measured context. The main challenges are threefold: spatial HRIR variation is globally correlated yet irregularly sampled; temporal responses contain high-frequency structure that is sensitive to small errors; and binaural consistency must be preserved to maintain physically meaningful interaural cues. Our design explicitly separates spatial reasoning from temporal reconstruction and introduces auxiliary cue supervision for stability under sparse measurements.

For each subject, measured and target directions are combined into a unified set of $L = M + N$ directions. We construct $\mathbf{H}_{\text{full}} \in \mathbb{R}^{L \times 2K}$
where measured rows contain observed HRIRs and target rows are initialized as zeros. A binary mask $\mathbf{i} \in \{0,1\}^{L}$ indicates observed directions, the $\ell$-th element $i_\ell$ indicates whether $\mathbf{h}_\ell$ is measured ($i_\ell=1$) or unmeasured ($i_\ell=0$). The network predicts HRIRs jointly for all directions, and measured rows are preserved via masked fusion.

\noindent
\textbf{Geometric Encoding:} Each binaural HRIR $\mathbf{h}_\ell \in \mathbb{R}^{1 \times 2K}$ is masked as $\tilde{\mathbf{h}}_\ell = i_\ell \mathbf{h}_\ell$ and projected into a latent feature: 
$\mathbf{e}_\ell = \mathrm{Norm}\left( \sigma(\tilde{\mathbf{h}}_\ell \mathbf{W}_s) \right)$, where $\sigma(\cdot)$ denotes GELU activation and $\mathrm{Norm}(\cdot)$ denotes layer normalization.

Each direction coordinate $\mathbf{x}_\ell \in \mathbb{R}^3$ is encoded using multi-frequency sinusoidal embedding~\cite{attention}
\begin{equation}
\label{eq:sinusoidal}
\begin{aligned}
\gamma(\mathbf{x}_\ell)=\big[
&\sin(2^0\pi \mathbf{x}_\ell),\cos(2^0\pi \mathbf{x}_\ell),\dots, \\
&\sin(2^{P-1}\pi \mathbf{x}_\ell),\cos(2^{P-1}\pi \mathbf{x}_\ell)
\big],
\end{aligned}
\end{equation}
with $P=6$. The geometric embedding is projected into the same latent space and added to the signal feature:
\begin{equation}
\mathbf{p}_\ell
=
\mathrm{Norm}\left(
\sigma([\mathbf{x}_\ell,\gamma(\mathbf{x}_\ell)] \mathbf{W}_g)
\right), \quad
\mathbf{o}_\ell
=
[\mathbf{e}_\ell,~ \mathbf{p}_\ell].
\end{equation}
Stacking all tokens $\mathbf{o}_\ell$ yields $\mathbf{O} \in \mathbb{R}^{L \times D}$.

\noindent
\textbf{Transformer Encoder:} The token sequence $\{\mathbf{o}_\ell\}_{\ell=1}^{L}$ is processed by a multi-layer Transformer encoder. Each layer consists of multi-head self-attention followed by a feed-forward network with residual connections and normalization.

Self-attention~\cite{attention} enables each direction to attend to all others, capturing global spatial dependencies across irregular directional layouts. The mask $\mathbf{i}$ is applied such that missing directions do not contribute as keys during attention. After $T$ layers, we obtain contextual representations: $\mathbf{C} \in \mathbb{R}^{L \times D}$.

\noindent
\textbf{Decoder and Conv1D Temporal Refinement:}  Each contextual feature $\mathbf{c}_\ell \in \mathbb{R}^{1 \times D}$ is mapped to a full-length binaural HRIR using a shared multi-layer perceptron (Fig.~\ref{fig:pipeline}):
\begin{equation}
\hat{\mathbf{h}}_\ell = f_{\text{dec}}(\mathbf{c}_\ell), \quad \hat{\mathbf{h}}_\ell \in \mathbb{R}^{1 \times 2K}.
\end{equation}

Stacking predictions $\hat{\mathbf{h}}_\ell$ yields $\hat{\mathbf{H}}_{\text{raw}} \in \mathbb{R}^{L \times 2K}$.

Measured rows are preserved through masked fusion:
\begin{equation}
\hat{\mathbf{H}}_{\text{fused}}
=
\mathbf{i} \odot \mathbf{H}_{\text{full}}
+
(1-\mathbf{i}) \odot \hat{\mathbf{H}}_{\text{raw}}.
\end{equation}

To improve local temporal coherence of the HRIRs, we apply a lightweight one-dimensional convolution. First, $\hat{\mathbf{H}}_{\text{fused}}$ is reordered by rows via elevation descending then azimuth ascending. Then, $\hat{\mathbf{H}} = \mathrm{Conv1D}(\hat{\mathbf{H}}_{\text{fused}})$, with residual blending controlled by $\mathbf{i}$ to ensure exact preservation of observed HRIRs.

\noindent
\textbf{Training Objective:} We supervise reconstruction only at missing directions, and define the reconstruction loss as:
\begin{equation}
\mathcal{L}_{\text{rec}}
=
\frac{1}{N}
\sum_{\ell: i_\ell = 0}
\left\lVert
\hat{\mathbf{h}}_\ell
-
\bar{\mathbf{h}}_\ell
\right\rVert_2^2.
\end{equation}

To improve spectral fidelity, we compute the orthonormally normalized discrete Fourier transform (scaled by $1/\sqrt{K}$), denoted by $\operatorname{DFT}(\cdot)$, for the left (L) and right (R) ears separately:
\begin{equation}
\mathcal{L}_{\text{HRTF}}
=
\sum_{c\in\{L,R\}}
\left\lVert
\operatorname{DFT}(\hat{\mathbf{H}}^{(c)})-\operatorname{DFT}(\bar{\mathbf{H}}^{(c)})
\right\rVert_2^2 .
\end{equation}

Auxiliary heads $f_{\text{head}}(\cdot)$ predict $\widehat{\mathrm{ITD}}_\ell$ and $\widehat{\mathrm{ILD}}_\ell$:
\begin{equation}
\big[ \widehat{\mathrm{ITD}}_\ell,~ \widehat{\mathrm{ILD}}_\ell \big] = 
f_{\text{head}}(\mathbf{c}_\ell),
\end{equation}

The corresponding ITD/ILD loss terms, $\mathcal{L}_{\text{ITD}}$ and $\mathcal{L}_{\text{ILD}}$, are:
\begin{equation}
\mathcal{L}_{\text{head}}
=
\frac{1}{N}\sum_{\ell:\, i_\ell=0}
\left\lVert
\widehat{\text{head}}_\ell-\text{head}_\ell
\right\rVert,
\quad
\text{head}\in\{\mathrm{ITD},\mathrm{ILD}\}.
\end{equation}

The total training loss $\mathcal{L}_{\text{total}}$ combines all components:
\begin{equation}
\mathcal{L}_{\text{total}}
=
\mathcal{L}_{\text{rec}}
+
\lambda_{\text{HRTF}} \mathcal{L}_{\text{HRTF}}
+
\lambda_{\text{ITD}} \mathcal{L}_{\text{ITD}}
+
\lambda_{\text{ILD}} \mathcal{L}_{\text{ILD}},
\end{equation}
where $\lambda_{\text{ITD}}$, $\lambda_{\text{ILD}}$, and $\lambda_{\text{HRTF}}$ are loss weights.


\begin{table*}[t]
  \centering
  \caption{ITD-E ($\mu$s) and ILD-E (dB) under different measurement sparsity levels $M$ on SONICOM. Bold indicates the best result per column. Baseline results are taken from~\cite{ranf}.}
  \vspace{-0.2cm}
  \label{tab:itd-ild}
  \footnotesize
  \setlength{\tabcolsep}{3.2pt}
  \renewcommand{\arraystretch}{1.05}
  \begin{tabular*}{\textwidth}{@{\extracolsep{\fill}}lcccccccc@{}}
    \toprule
    \multirow{2}{*}{\textbf{Method}}
      & \multicolumn{2}{c}{$M=3$}
      & \multicolumn{2}{c}{$M=5$}
      & \multicolumn{2}{c}{$M=19$}
      & \multicolumn{2}{c}{$M=100$} \\
    \cmidrule(lr){2-3}\cmidrule(lr){4-5}\cmidrule(lr){6-7}\cmidrule(lr){8-9}
      & \makecell[c]{ITD-E~($\mu$s)$\downarrow$}
      & \makecell[c]{ILD-E~(dB)$\downarrow$}
      & \makecell[c]{ITD-E~($\mu$s)$\downarrow$}
      & \makecell[c]{ILD-E~(dB)$\downarrow$}
      & \makecell[c]{ITD-E~($\mu$s)$\downarrow$}
      & \makecell[c]{ILD-E~(dB)$\downarrow$}
      & \makecell[c]{ITD-E~($\mu$s)$\downarrow$}
      & \makecell[c]{ILD-E~(dB)$\downarrow$} \\
    \midrule
    \textbf{\textit{HRIR-Former}} (Proposed)
      & \textbf{18.5} & \textbf{1.14}
      & \textbf{16.4} & \textbf{1.10}
      & 15.4 & \textbf{0.95}
      & 10.6 & \textbf{0.70} \\
    \midrule
    Nbr~\cite{ranf}
      & 274.2 & 7.6 & 154.2 & 4.8 & 108.3 & 3.0 & 44.0 & 1.4 \\
    HRTF-Sel-ITD~\cite{ranf}
      & 26.3 & 1.4 & 24.5 & 1.5 & 23.2 & 1.6 & 20.0 & 1.4 \\
    HRTF-Sel-LSD~\cite{ranf}
      & 37.5 & 1.5 & 35.7 & 1.4 & 37.4 & 1.5 & 31.4 & 1.3 \\
    \midrule
    NF-CbC~\cite{NF-CbC}
      & 22.1 & 1.5 & 20.7 & 2.0 & 14.8 & 1.8 & 11.8 & 1.7 \\
    NF-LoRA~\cite{NF-LoRA}
      & 28.6 & 1.3 & 24.7 & 1.4 & 14.7 & 1.1 & \textbf{9.1} & 1.1 \\
    RANF~\cite{ranf}
      & 20.5 & 1.2 & 18.7 & 1.2 & \textbf{14.2} & 1.0 & 10.0 & 0.8 \\
    \bottomrule
  \end{tabular*}
  \vspace{-0.3cm}
\end{table*}



\section{Experiment and Validation}
\label{sec:experiments}

\subsection{Experimental Setup}
\label{ssec:experimental_setup}

We evaluate the binaural HRIR spatial up-sampling performance of \textit{HRIR-Former} on the SONICOM database~\cite{sonicom-data}, which contains measurements for over 200 subjects with 793 source directions per subject. Each HRIR is sampled at 48~kHz and represented with $K=256$ samples. For a fair comparison, we follow the protocol of~\cite{ranf}: the first 180 subjects are used for training and the subsequent 20 subjects are used for validation; subject \texttt{P0079} is excluded due to atypical ITD behavior~\cite{ranf}. Since \textit{HRIR-Former} reconstructs HRIRs in the time domain and does not impose a MP assumption, we train and evaluate it on SONICOM free-field compensated HRIRs (no MP filtering applied)~\cite{sonicom-data}. During training, we set the loss weights $\lambda_{\text{ITD}}=\lambda_{\text{ILD}}=0.05$ and $\lambda_{\text{HRTF}}=500$. 
The encoder uses $T{=}3$ layers with model dimension $D{=}256$ and $4$ attention heads. The Conv1D temporal refinement applies a kernel of size~3, smoothing across spatially adjacent directions.
We train \textit{HRIR-Former} using AdamW~\cite{adamW} with learning rate $3\times10^{-4}$ and batch size 8 for 500 epochs on an NVIDIA A100 GPU.


\vspace{-0.15cm}
\subsection{Evaluation Metrics}
\label{ssec:evaluation_metrics}

To quantify binaural HRIR spatial up-sampling performance, we report four metrics: (1) ITD Error (ITD-E)~\cite{LAPreport}, (2) ILD Error (ILD-E)~\cite{LAPreport}, (3) Normalized Mean Squared Error (NMSE)~\cite{rir-former}, and (4) Cosine Distance (CD)~\cite{rir-former,CD,DiffusionRIR}. Formal definitions follow the referenced works. NMSE and CD evaluate time-domain waveform fidelity and shape alignment between the estimated and ground-truth HRIRs; CD measures scale-invariant similarity and ranges in $[0,2]$, where $\text{CD}=0$ indicates identical waveform shapes~\cite{rir-former}. ITD-E is the mean absolute error between reconstructed and ground-truth ITDs, reflecting binaural timing discrepancies relevant to localization~\cite{LAPreport}. ILD-E measures errors in interaural level differences across frequency and direction; larger ILD-E may lead to lateralization shifts or degraded spatial clarity~\cite{LAPreport}. Lower values indicate better performance for all metrics. All metrics are averaged over $N$ target directions and subjects.


\vspace{-0.15cm}
\subsection{Comparison Methods}
\label{ssec:comparison_methods}

We compare \textit{HRIR-Former} with six baselines: Nearest Neighbor (Nbr), HRTF-Sel-ITD, HRTF-Sel-LSD~\cite{ranf}, NF-CbC~\cite{NF-CbC}, NF-LoRA~\cite{NF-LoRA}, and RANF~\cite{ranf}. Since these methods primarily reconstruct HRTF magnitude responses in the frequency domain, we focus on the shared binaural cues, ITD-E and ILD-E, for direct comparison with our time-domain approach. As we adopt the same training/evaluation protocol as~\cite{ranf}, we use the ITD-E and ILD-E results reported in~\cite{ranf} for these baselines and summarize them in Table~\ref{tab:itd-ild}.


\vspace{-0.15cm}
\subsection{Results}
\label{ssec:results}

Table~\ref{tab:itd-ild} reports ITD-E and ILD-E under four measurement sparsity levels ($M=3,5,19,100$). \textit{HRIR-Former} achieves the lowest ILD-E across all $M$, indicating consistently accurate interaural level cues, and attains the best ITD-E in the most sparse regimes ($M=3$ and $M=5$). For denser inputs ($M=19$ and $M=100$), \textit{HRIR-Former} remains competitive, with ITD-E comparable to the strongest frequency-domain baselines. 
In contrast, the Nbr baseline yields substantially larger ITD-E and ILD-E across all $M$, highlighting the difficulty of spatial up-sampling from sparse measurements and the benefit of learned reconstruction~\cite{ranf}.

\begin{figure}[t]
  \centering
  \includegraphics[width=1\linewidth]{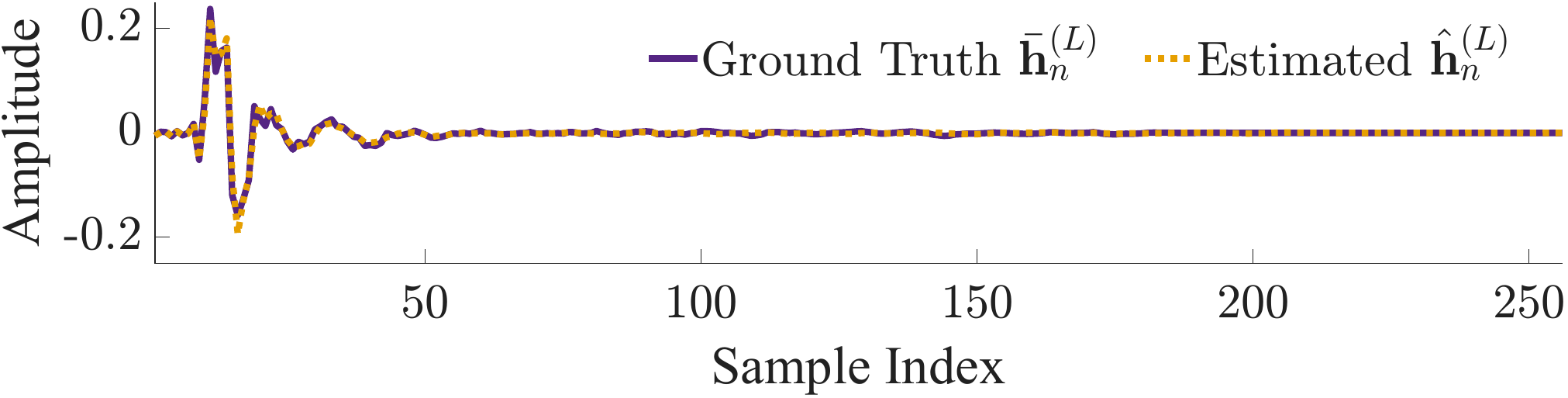}
  \vspace{-0.45cm}
  \caption{Left-ear HRIR reconstruction for subject \texttt{P0187} at $(\phi,\theta,r)=(90^\circ,20^\circ,1.5~\mathrm{m})$ under $M=19$: estimated HRIR (dashed) versus ground-truth HRIR (solid). Angular distance to the nearest measured direction: $35.2^\circ$.}
  \label{fig:hrir-line}
  \vspace{-0.45cm}
\end{figure}
\begin{figure}[t]
  \centering
  \includegraphics[width=1\linewidth]{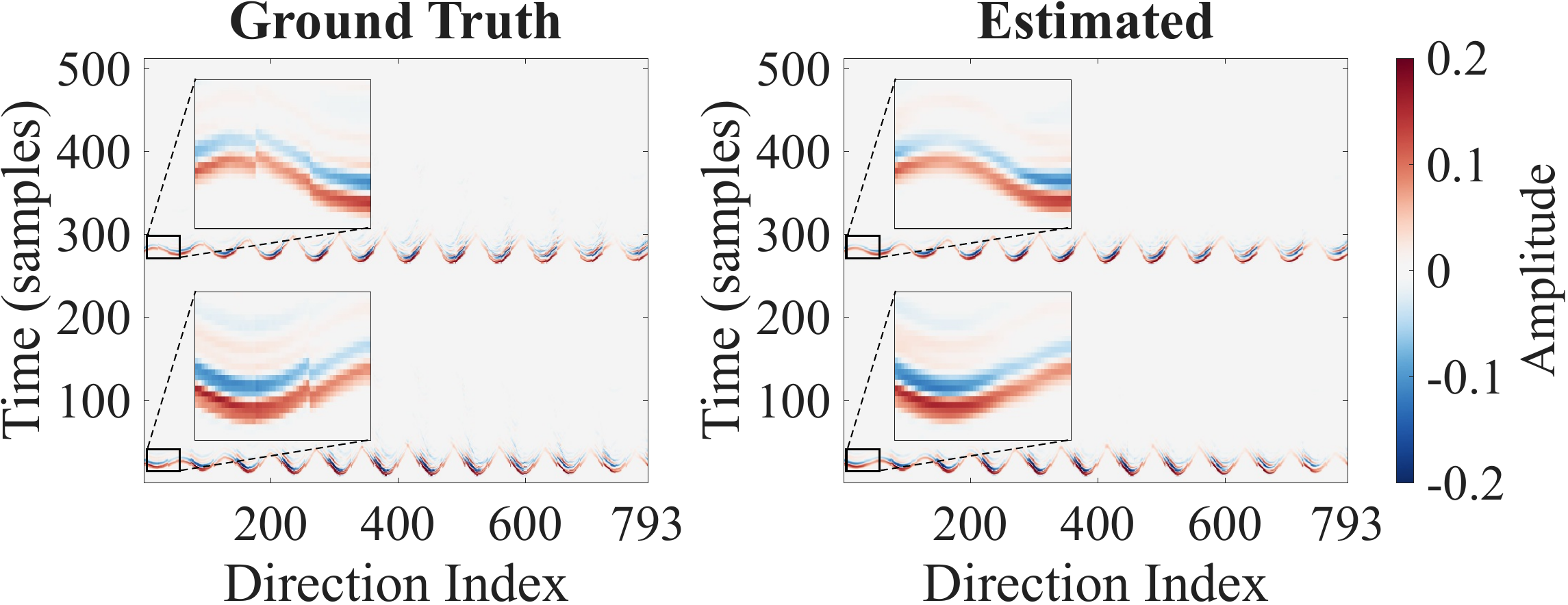}
  \caption{Stacked binaural HRIRs for subject \texttt{P0187} under $M=19$ over 793 directions (each column: one direction; left/right ears concatenated along time). Ground truth (left) and estimation (right).}
  \label{fig:hrir-2d}
  \vspace{-0.5cm}
\end{figure}

Table~\ref{tab:cd_nmse} summarizes time-domain reconstruction accuracy. Across all $M$, \textit{HRIR-Former} achieves NMSE below $-6.90$~dB and CD below $0.233$, and both metrics improve as $M$ increases. Fig.~\ref{fig:hrir-2d} visualizes stacked binaural HRIRs across all 793 directions, showing that the estimated HRIR field closely matches the ground truth globally. Fig.~\ref{fig:hrir-line} shows an example (left ear) under $M=19$, where the predicted waveform closely matches the ground truth in both onset timing and fine structure.


\begin{table}[t]
  \centering
  \caption{NMSE and CD of \textit{HRIR-Former} under different measurement sparsity levels $M$.}
  \vspace{-0.15cm}
  \label{tab:cd_nmse}
  \footnotesize
  \setlength{\tabcolsep}{2.5pt}
  \renewcommand{\arraystretch}{0.95}
  \begin{tabular*}{\columnwidth}{@{\extracolsep{\fill}}lcccc@{}}
    \toprule
    \textbf{Metric} & \textbf{$M{=}3$} & \textbf{$M{=}5$} & \textbf{$M{=}19$} & \textbf{$M{=}100$} \\
    \midrule
    NMSE (dB)$\downarrow$ & -6.90 & -7.35 & -8.24 & \textbf{-10.20} \\
    CD$\downarrow$        & 0.233 & 0.210 & 0.163 & \textbf{0.102}  \\
    \bottomrule
  \end{tabular*}
\end{table}


\begin{table}[ht]
  \centering
  \caption{Ablation study of \textit{HRIR-Former} under $M=5$.}
  \vspace{-0.3cm}
  \label{tab:ab}
  \footnotesize
  \setlength{\tabcolsep}{3.5pt}
  \renewcommand{\arraystretch}{1.08}
  \begin{tabular}{>{\raggedright\arraybackslash}p{0.46\columnwidth}cccc}
    \toprule
    \textbf{Variant}
    & \makecell[c]{\textbf{NMSE}\\\textbf{(dB)}~$\downarrow$}
    & \makecell[c]{\textbf{CD}~$\downarrow$}
    & \makecell[c]{\textbf{ITD-E}\\\textbf{($\mu$s)}~$\downarrow$}
    & \makecell[c]{\textbf{ILD-E}\\\textbf{(dB)}~$\downarrow$} \\
    \midrule
    \textit{HRIR-Former} (full)
      & -7.35 & 0.210 & \textbf{16.4} & \textbf{1.10} \\
    \midrule
    w/o sinusoidal encoding
      & -4.99 & 0.298 & 24.2 & 1.66 \\
    w/o ITD/ILD-prediction heads
      & \textbf{-7.52} & \textbf{0.207} & 19.4 & 1.11 \\
    w/o Conv1D refinement
      & -6.04 & 0.257 & 20.8 & 1.28 \\
    w/o $\mathcal{L}_{\text{HRTF}}$
      & -7.10 & 0.217 & 17.3 & 1.14 \\
    w/ MP pre-processing
      & -6.80 & 0.225 & 19.1 & 1.21 \\
    \bottomrule
  \end{tabular}
  \vspace{-0.3cm}
\end{table}


\subsection{Ablation Study}
\label{ssec:ablation}

We conduct ablation studies under the $M=5$ setting: (1) removing the sinusoidal direction-encoding module in Eq.~\ref{eq:sinusoidal}, (2) removing the auxiliary ITD/ILD-prediction heads from the network outputs, (3) removing the post-Transformer Conv1D temporal refinement module, (4) removing the complex HRTF loss $\mathcal{L}_{\text{HRTF}}$, and (5) training and testing on MP-preprocessed HRIRs. Table~\ref{tab:ab} summarizes the results. Removing any component degrades performance on multiple metrics, indicating that each design choice contributes to reconstruction quality.

Removing the sinusoidal encoding causes the largest degradation across NMSE, CD, ITD-E, and ILD-E, highlighting the importance of rich geometric conditioning. In particular, sinusoidal encoding provides a continuous representation of $\mathbf{x}^{\text{tgt}}_n=(\phi,\theta,r)$, enabling grid-free inference at previously unseen directions. Training and testing on MP-preprocessed HRIRs degrades performance versus the default setting, indicating that \textit{HRIR-Former} does not impose an MP assumption and does not benefit from MP pre-processing. The ITD/ILD-prediction heads primarily improve binaural-cue accuracy, while having a marginal effect on NMSE and CD. Finally, removing the Conv1D temporal refinement degrades all metrics, consistent with its role in improving local consistency across neighboring directions; as illustrated by the zoom-in region in Fig.~\ref{fig:hrir-2d}, this refinement mitigates local discontinuities (e.g., from small measurement inconsistencies such as slight head motion) that could otherwise manifest as audible clicks during binaural rendering.


\section{Conclusion}

We presented \textit{HRIR-Former}, a time-domain, grid-free binaural Transformer for continuous angular (azimuth--elevation) HRIR spatial up-sampling, reconstructing HRIRs at arbitrary target directions from sparse per-listener measurements. \textit{HRIR-Former} combines sinusoidal direction encoding, auxiliary ITD/ILD prediction heads, a post-Transformer Conv1D temporal refinement module, and a complex HRTF loss $\mathcal{L}_{\text{HRTF}}$. Experiments on SONICOM free-field HRIRs show strong performance on time-domain metrics (NMSE, CD) and binaural-cue metrics (ITD-E, ILD-E). Ablations confirm the contribution of each component and suggest that explicit MP pre-processing is not required. Future work will include deeper model analysis and perceptual listening tests for binaural rendering.


\section{Acknowledgments}

Shaoheng Xu and Chunyi Sun are each supported by an ANU PhD Scholarship and an ANU HDR Fee Merit Scholarship from The Australian National University.


\section{Generative AI Use Disclosure}

We used ChatGPT (OpenAI) for English-language editing and stylistic polishing. The tool was not used to generate technical content, experimental results, or conclusions. All text was reviewed and approved by the authors.


\bibliographystyle{IEEEtran}
\bibliography{mybib}


\end{document}